\documentclass [12 pt]{article}
\usepackage{psfig}
\usepackage{epsfig}
\begin{document}
\title { A reformulation of intrabeam scattering theory}
\author{George Parzen}
\date{April 30, 2004 \\BNL Report C-A/AP/No. 150}
\maketitle
\begin{abstract}
The motivation for the  treatment of intrabeam scattering theory 
given in this paper was to find
results which  would be convenient for  computing the 
intrabeam scattering growth rates for particle distributions which are more
complicated than a simple gaussian.
It was shown by A. Piwinski that
beam growth rates due to intrabeam scattering can be expressed 
as a multidimensional  integral [1]. 
It was pointed out by J. Bjorken and S. Mtingwa [2] 
that the 
reduction of the multidimensional integral to a 3-dimensional untegral
is made easier by 
writing the integral so that its 
relativistic transformation properties are more obvious.
The starting point in [2] was a result from 
the treatment of the  
two body scattering problem in relativistic quantum theory .  
In this paper the starting point is the relativistic transformation 
properties of the scattering cross section which may be a 
more familiar starting point. 
The resulting expression for the multidimensional integral is simpler to reduce.
In addition, the results do not depend on the particular form of 
the Coulomb cross section that was used in [2] and are valid for any collision 
cross section.
\end{abstract}

\section{Introduction}

The motivation for the  treatment of intrabeam scattering theory 
given in this paper was to find
results which  would be convenient for  computing the 
intrabeam scattering growth rates for particle distributions which are more
complicated than a simple gaussian.
It was shown by A. Piwinski that
beam growth rates due to intrabeam scattering can be expressed 
as a multidimensional  integral [1]. 
It was pointed out by J. Bjorken and S. Mtingwa [2] 
that the 
reduction of the multidimensional integral to a 3-dimensional untegral
is made easier by 
writing the integral so that its 
relativistic transformation properties are more obvious.
The starting point in [2] was a result from 
the treatment of the  
two body scattering problem in relativistic quantum theory .  
In this paper the starting point is the relativistic transformation 
properties of the scattering cross section which may be a 
more familiar starting point. 
The resulting expression for the multidimensional integral is simpler to reduce.
In addition, the results do not depend on the particular form of 
the Coulomb cross section that was used in [2] and are valid for any collision 
cross section.The final result is given by Eq.(14), which can be 
used for  computing the 
intrabeam scattering growth rates for particle distributions which are more
complicated than a simple gaussian.

\section{Transformation properties of the cross section}

The cross section, $\sigma$, which describes the the scattering of particles 
with the momentum $p_1$ from the target particles with momentum $p_2$  is first 
defined in the CS (coordinate system) where the target particles are at rest,
$p_2=0$. In a scattering event, the particle momenta change from 
$p_1,p_2$ to $p_1',p_2'$. As we are assuming that both momentum and energy are
conserved , the final momenta, $p_1',p_2'$ are determined by the direction of 
$p_1'$ which is indicated by the unit vector $\hat{p_1'}$. In this CS where
$p_2=0$, $\sigma$ is defined so that the number of incident particles 
which are scattered by the target particles  with momemtum $p_2$ which are in the 
volume element, $d^3x$, in the time interval $dt$, into the 
solid angle $d\Omega'$ corresponding to the
direction $\hat{p_1'}$ is given by
\begin{eqnarray} 
\delta N &=& \sigma d\Omega' \rho_1(x) v_1 \rho_2(x) d^3x dt
\end{eqnarray}
where $\rho_1(x),\rho_2(x)$ are the density functions and $v_1$ is the 
velocity of the incident particle. 

Now let us go to a CS where $p_2 \neq 0$.
In this CS, $\sigma$ is defined by requiring $\sigma d\Omega'$ to be invariant, 
that is to have the same value in all coordinate systems.
A simple way to find the relationship between $\delta N$ and  $\sigma$
in this CS is to write $\delta N$ as (see [3])
\begin{eqnarray} 
\delta N &=& \sigma d\Omega' \frac{\rho_1(x)} {\gamma_1}
      \frac{\rho_2(x)}{\gamma_2}  F(p_1,p_2)d^3x dt
\end{eqnarray}
where $\rho_1(x) / \gamma_1$ is an invariant as it is just  the 
density function for particle 1 in the CS where $p_1=0$. Similarly for
$\rho_2(x)/\gamma_2$. If one can find an invariant $F(p_1,p_2)$ which for 
$p_2=0$ gives $F=\gamma_1 v_1$, then this expression for $\delta N$ gives the
correct result when $p_2=0$ and also gives the correct result 
when $p_2 \neq 0$. $F(p_1,p_2)$ that satifies these requirements is
\begin{eqnarray}
F(p_1,p_2) &=& c \frac {[(p_1p_2)^2-m_1^2m_2^2c^4]^{.5}}{m_1m_2 c^2}
\end{eqnarray}
Here, $p_1,p_2$ are 4-vectors whose first three components are the components
of the momemtum and the fourth component is  $ iE/c $, $E=(p^2c^2+m^2c^4)^{.5}$.
$F(p_1,p_2)$ is an invariant and when $p_2=0$, $F=\gamma_1 v_1$.
The result for $F(p_1,p_2)$ given by Eq.(3) can also be written as
\begin{eqnarray}
F(p_1,p_2) &=& \gamma_1 \gamma_2 c [(\vec{\beta_1}-\vec{\beta_2})^2 -(\vec{\beta_1} \times \vec{\beta_2})^2]^{.5}
\end{eqnarray}
Here, $\vec{\beta_1},\vec{\beta_2}$ are vectors in 3-space corresponding to 
the velocities of the particles divide by $c$.

\section{The $f(x,p)$ distribution and the scattering rate $\delta N$}

Let us now treat the case where the particles are contained within a bunch 
and their distibution is given by $f(x,p)$ where $N f(x,p)$ is the number 
of particles in $d^3xd^3p$. N is the number of particles in the bunch, 
all particles have the same rest mass $m$ and
\[ \int d^3xd^3p \; f(x,p)=1  \]
Let $\delta N$ be the number of  particles with momentum $p_1$ in $d^3p_1$ 
and  space coordinate $x$ in $d^3x$
which are scattered by the particles  with momentum $p_2$ in $d^3p_2$ 
which are also in  $d^3x$, in the time interval $dt$ , into the 
solid angle $d\Omega'$ corresponding to the
direction $\hat{p_1'}$. Then $\delta N$ can be obtained 
 using the same procedure used in 
obtaining Eq.(2), provided one knows 
that $d^3p/\gamma $ and  $f(x,p)$ are
invariants, which is shown in section 5. $\delta N$ is given by
\begin{eqnarray} 
\delta N &=& N^2 \sigma d\Omega' \frac {d^3p_1}{\gamma_1}
      \frac {d^3p_2}{\gamma_2} f(x,p_1)f(x,p_2) F(p_1,p_2) d^3x dt \nonumber\\
F(p_1,p_2) &=&  \frac {[(p_1p_2)^2-m^4 ]^{.5}}{m^2}
\end{eqnarray}
One may note that the right hand side of this expression for $\delta N$ is
an invariant. We will be putting $c=1$ except when something may be gained
by showing $c$ explicitly.

\section{Growth rates for $<p_ip_j>$}

Growth rates will be  given for $<p_i p_j>$. where the $<>$ indicate an 
average over all the particles in the bunch. From these one can compute the 
growth rates for the emittances, $<\epsilon_i>$. The advantage due to 
computing growth rates for $<p_i p_j>$ stems from the observation that if  
$p_i,p_j$ are the components of the momentum 4-vector, then $p_ip_j$
is a tensor in 4-space and so is $\delta <p_ip_j>$, as will be seen below, 
where $\delta <p_ip_j>$ is the change in $<p_ip_j>$ in a time interval $dt$.
The transfornation properties of a tensor can then be used to facilitate the
transfer of results between two CS.

In a scattering event, where a particle with  momentum $p_1$ scatters off a 
particle with momentum $p_2$, the momenta will change to $p_1'$ and $p_2'$.
Let $\delta p_{1i}$ represent the change in $p_{1i}$ in the collision, 
and similarly for $\delta (p_{1i}p_{1j})$. Then
\begin{eqnarray} 
\delta p_{1i} &=& p_{1i}'-p_{1i}     \nonumber\\
\delta (p_{1i}p_{1j}) &=& p_{1i}' p_{1j}'-p_{1i}p_{1j}
\end{eqnarray}

Using the scattering rate given by Eq.(5), one can now compute 
$\delta<p_ip_j>$
\begin{eqnarray} 
<\delta (p_{1i}p_{1j}) > &=& N \int \;\: d^3x \frac {d^3p_1}{\gamma_1}
      \frac {d^3p_2}{\gamma_2} f(x,p_1)f(x,p_2) F(p_1,p_2)   \nonumber\\
      & &  \sigma d\Omega'(p_{1i}' p_{1j}'-p_{1i}p_{1j}) dt   \nonumber\\
F(p_1,p_2) &=&  \frac {[(p_1p_2)^2-m^4 ]^{.5}}{m^2}
\end{eqnarray}
One may note that 
\[<\delta (p_{1i}p_{1j}) >=\delta <p_{1i}p_{1j} > \]
and that $\delta <p_{1i}p_{1j} >$ is a tensor in 4-space because of the 
transformation properties given above for the quantities appearing on the 
right hand side of  Eq.(7). Eq.(7) is our general result for the growth 
rates , holds in all CS, and can be used for any particle distribution,
$f(x,p)$.

This result can be further simplified by first
considering the integral, for a given $p_1,p_2$,
\begin{eqnarray} 
C_{ij} &=& \int \sigma d\Omega'(p_{1i}' p_{1j}'-p_{1i}p_{1j})    
\end{eqnarray}
$C_{ij}$ has the transformation properties of a tensor in 4-space as
$\sigma d\Omega'$ is an invariant. For a given $p_1,p_2$, $C_{ij}$ 
can be evaluated in the CMS ( the center of mass CS ) and if the result
can be written in terms of 4-vectors and tensors in 4-space, then the result
in this form. will hold in all CS. The calculation of $C_{ij}$ can be simplified
by noting that because of the symmetry in $p_1$ and $p_2$ we have
\begin{eqnarray} 
<\delta (p_{1i}p_{1j})> &=&  <\delta (p_{2i}p_{2j})>
\end{eqnarray}
and we can define $C_{ij}$ as
\begin{eqnarray} 
C_{ij} &=& \int \sigma d\Omega'\frac{1}{2} [\delta (p_{1i}p_{1j})+\delta (p_{2i}p_{2j})]    
\end{eqnarray}
and Eq.(7) can be written as
\begin{eqnarray} 
<\delta (p_{1i}p_{1j}) > &=& N \int \;\: d^3x \frac {d^3p_1}{\gamma_1}
      \frac {d^3p_2}{\gamma_2} f(x,p_1)f(x,p_2) F(p_1,p_2)  C_{ij} dt
\end{eqnarray}

We will now further evaluate $C_{ij}$ by first evaluating $C_{ij}$  
for some particular values of $p_1$,$p_2$ in the CMS corresponding to
$p_1$,$p_2$ and and then using the tensor properties of $C_{ij}$ 
to find a result that holds in any other CS. We are particularly interested
in finding a result in the Rest CS, which is the CS which moves 
along with the bunch. In the CMS, 
\begin{eqnarray*} 
\vec{p_2} &=& -\vec{p_1}   \nonumber\\
\vec{\Delta} &=& \frac{1}{2} (\vec{p_1}-\vec{p_2})=\vec{p_1}  \nonumber\\
\vec{q_1} &=& \vec{p_1'}-\vec{p_1} \nonumber\\
\vec{q_2} &=& -\vec{q_1}
\end{eqnarray*}
Using $\vec{q_1} = \vec{p_1'}-\vec{p_1}$ and $\vec{q_2} = -\vec{q_1}$,
one can show that
\begin{eqnarray} 
\frac{1}{2} (\delta (p_{1i}p_{1j})+\delta (p_{2i}p_{2j})  &=& 
          q_{1i} q_{1j}+\Delta_i q_{1j}+ \Delta_j q_{1i}
\end{eqnarray}

In the CMS, we introduce a polar coordinate system $\theta,\phi$
where $\theta$ is measured relative to the direction of 
$\vec{p_1}$ or $\vec{\Delta}$ 
and we assume that $\sigma(\theta,\phi)$ is a fumction of $\theta$ only.
we can then write
\begin{eqnarray*} 
\vec{p_1} &=& (0,0,1)|\vec{\Delta}|   \nonumber\\
\vec{p_1'} &=& (\sin \theta \cos \phi,\sin \theta \sin \phi,
          \cos \theta)|\vec{\Delta}|    \nonumber\\
\vec{q_1} &=& (\sin \theta \cos \phi,\sin \theta \sin \phi,
          \cos \theta-1)|\vec{\Delta}| 
\end{eqnarray*}
Considering $p_1$,$p_2$ to be 4-vectors, and $\Delta=.5(p_1-p_2)$,
$q_1=p_1'-p_1$, then $\Delta$, $q_1$ are also 4-vectors and
in the CMS, $\Delta_4=0,q_{14}=0$. 

Using Eqs(10) and (12), one now finds for $C_{ij}$ in the CMS
\[ C_{ij} = \pi \int_{0}^{\pi} d\theta \sigma \sin^3 \theta \;|\vec{\Delta}|^2
            \left( \begin{array}{ccrc} 1&0&0&0    \\
                                       0&1&0&0    \\
                                       0&0&-2&0    \\
                                       0&0&0&0   
                                      \end{array} \right )   \]
To find $C_{ij}$ in the Rest CS or in the Lab CS, we will try to 
find an expression for $C_{ij}$ in terms of the 4-vectors 
$p_{1i},p_{2i}$  which gives the above result for $C_{ij}$ in the
CMS. The expression that does this is given by
\begin{eqnarray} 
C_{ij} &=& \pi \int_{0}^{\pi} d\theta \sigma \sin^3 \theta \;
       \Delta^2 [\delta_{ij}-3\frac{\Delta_i \Delta_j}{\Delta^2} 
       +\frac{W_iW_j}{W^2}] \;\; i,j=1,4   \nonumber\\
\Delta_i &=& \frac{1}{2}(p_{1i}-p_{2i})   \nonumber\\
W_i &=& p_{1i}+p_{2i}
\end{eqnarray}
where $\sigma$ is the cross section in the CMS. Let us now verify that 
this expression gives the correct result for $C_{ij}$ in the CMS.
In the CMS,
\begin{eqnarray*} 
\Delta^2 &=& |\vec{\Delta}|^2 +\Delta_4^2=|\vec{\Delta}|^2   \nonumber\\
W_i &=& 0 \;\; i=1,3                                         \nonumber\\
\Delta_i &=& 0 \;\; i=1,2 \;\;\Delta_3=|\vec{\Delta}|
\end{eqnarray*}
so that Eq.(13) does give the correct result in the CMS.

An important further simplification results from the fact that 
the particle motion is non-relativistic in the CMS and also in the Rest CS
which moves along with the bunch. For RHIC parameters, for $\gamma=100$,
one finds that $p \simeq 1e-3\;\;mc$. One can then drop the 
$W_iW_j/W^2$ term. Also $\Delta^2=|\vec{\Delta}|^2$ in the CMS and in the 
Rest CS and one can evaluate $F(p_1,p_2)$ using Eq.(4) as 
$F(p_1,p_2)=2 c \bar{\beta}$ where $\bar{\beta} c$ is the velocity of either
particle in the CMS . In the Rest CS, one can write
\begin{eqnarray} 
C_{ij} &=& \pi \int_{0}^{\pi} d\theta \sigma \sin^3 \theta \;
        [|\vec{\Delta}|^2 \delta_{ij}-3\Delta_i \Delta_j
      ] \;\; i,j=1,3   \nonumber\\
\Delta_i &=& \frac{1}{2}(p_{1i}-p_{2i})   \nonumber\\
<\delta (p_{1i}p_{1j}) > &=& N \int \;\: d^3x d^3p_1
      d^3p_2 f(x,p_1)f(x,p_2) 2 \bar{\beta}c \; C_{ij}\;dt  
      \nonumber\\
\bar{\beta}c &=& |\vec{\Delta}|/m
\end{eqnarray}

Eq.(14) would be a good starting point for computing growth rates for a 
particle distribution more complicated than a simple gaussian.
For the case of the Coulomb cross section, one can write $C_{ij}$ as
\begin{eqnarray} 
C_{ij} &=& 2 \pi (r_0/2 \bar{\beta}^2)^2 
      \ln(1+(2 \bar{\beta}^2 b_{max}/r_0)^2) \;\;  
       [|\vec{\Delta}|^2 \delta_{ij}-3\Delta_i \Delta_j
       ] \;\; i,j=1,3                             \nonumber\\
\sigma &=& (r_0/2 \bar{\beta}^2)^2/(1-\cos \theta )^2    \nonumber\\
r_0 &=& Z^2e^2/mc^2
\end{eqnarray}
$b_{max}$ is the largest allowed impact parameters in the CMS.

\section {Invariants $d^3p/\gamma$ and $f(x,p)$}

In order to eatablish Eq.(5), one needs to know that $d^3p/\gamma$ and $f(x,p)$
are invariants. Consider a CS moving with the velocity $v_0$ with respect 
to the Laboratory CS. Let the coordinates be $x,p$ in the Laboratory CS and 
$\hat{x},\hat{p}$ in the new CS. $\; p,\hat{p}$ are related by
\begin{eqnarray}
\hat{p_s} &=& \gamma_0(p_s-v_0 E)     \nonumber\\
\hat{p_x} &=& p_x               \nonumber\\
\hat{p_y} &=& p_y               \nonumber\\
E &=& \sqrt{p^2+m^2}          \nonumber\\
\gamma_0 &=& 1/\sqrt{1-v_0^2}
\end{eqnarray}
It then follows that
\begin{eqnarray}
d \hat{p_s} &=& \gamma_0(d p_s-v_0 dE)     \nonumber\\
dE &=& (p_s/E) dp_s \;\;\; p_x,p_y constant   \nonumber\\
d \hat{p_s} &=& \gamma_0(1-v_0 p_s/E)dp_s     \nonumber\\
  & &                                         \nonumber\\
\hat{E} &=& \sqrt{\hat{p}^2+m^2}               \nonumber\\
\hat{E} &=& \gamma_0(E-v_0 p_s)               \nonumber\\
\gamma_0(1-v_0 p_s/E) &=&  \hat{E}/E          \nonumber\\
\frac{d\hat{p_s}}{\hat{E}} &=& \frac{dp_s}{E}
\end{eqnarray}
Thus $dp_s/\gamma$ is invariant under this momentum transformation and also
$d^3p/\gamma\;$ is invariant.

Now let us show that $f(x,p)$ is an invariant. Since $f(x,p)d^3xd^3p$ 
is an invariant, as it gives the number of particles in $d^3xd^3p$, we need to show
that $d^3xd^3p$ is an invariant. Consider the point $x,p$ in some
CS. In the moving CS where $\hat{p_s}=0$, which is moving with the velocity
$v=p_s/E$ with respect to the first CS,
\begin{eqnarray}
d \hat{p_s} &=& \gamma(d p_s-v dE)     \nonumber\\
dE &=& (p_s/E) dp_s \;\;\; p_x,p_y constant   \nonumber\\
d \hat{p_s} &=& dp_s/\gamma                 \nonumber\\
\gamma &=& 1/\sqrt{1-v^2}
\end{eqnarray}
Since $dp_s/\gamma=d\hat{p_s}$ holds for any CS, $dp_s/\gamma$ is an
invariant and $d^3p/\gamma$ is an invariant. One also has 
$\gamma d^3x$ is  invariant because of the Lorentz-Fitsgerald contraction. 
We have then that that $d^3xd^3p$ is an invariant.

\section*{References}

 1. A. Piwinski Proc. 9th Int. Conf. on High Energy Accelerators (1974) 405,
   M. Martini CERN PS/84-9 (1984), A. Piwinski Cern 87-03 (1987) 402,
   A. Piwinski CERN 92-01 (1992) 226

2. J.D. Bjorken and S.K. Mtingwa, Part. Accel.13 (1983) 115, K. Kubo and
   K. Oide  Phys. Rev. S.T.A.B., 4, (2001) 124401

\end{document}